\documentstyle[aps,preprint]{revtex}

\begin{document}

\title{Reply to Comment on ``Dynamics of Weak First Order Phase Transitions''}
\author{Marcelo~Gleiser}
\address{Department of Physics and Astronomy\\
Dartmouth College\\
Hanover, NH 03755, USA}

\date{DART-HEP-95/02 $~~~~~~$ \today}
\maketitle
\pacs{PACS numbers: 98.80, 64.60 }

In the preceding Comment \cite{HJ}, Harris and Jungman (HJ) offer interesting
criticism to the interpretation I advanced in the Letter titled ``Dynamics
of weak first order phase transitions'' \cite{MG}. In order to clarify my
position, it is worthwhile spending a few sentences restating both my
results and HJ's comments in a language which may bridge most of our
differences.

The aim of this work was to study numerically
the role thermal fluctuations play in
promoting phase mixing in a system described by a Ginzburg-Landau free-energy
density with two minima which are degenerate at the critical
temperature $T_c$. Below $T_c$ one minimum has lower free energy and thus
becomes thermodynamically preferred. Let's call this free energy the
{\it mean field} free energy. This mean field free energy has an
implicit coarse-graining scale, the mean-field correlation length, which
is its inverse curvature. In the simulations,
the lattice spacing was chosen to be smaller
than the correlation length at all temperatures.
The idea was to prepare the system
initially in the minimum which is metastable below $T_c$ and measure the
fraction of the total area (the simulation was done in 2d) which remains in
this phase as the system evolves according to a Markovian Langevin
equation. The simulations were done at $T_c$, and the parameter which I
chose to vary was the coefficient of the cubic coupling, $\alpha$,
which measures the
strength of the {\it mean field} thermodynamic barrier between the two phases.
I found that below a certain value of $\alpha$, called $\alpha_c$, the
two phases mixed completely, while for $\alpha > \alpha_c$ the system
remained well-localized in the initial phase. Thus, even though the
{\it mean field} free-energy density had a barrier separating the two
phases for $\alpha < \alpha_c$, the system is better described by a
free-energy density with only one minimum, located at the top of the barrier
between the two phases. In other words, the thermal fluctuations washed away
the first-order phase transition.

At $\alpha_c$ I observed that the system
exhibited critical slowing down, as it should in the vicinity of a second
order phase transition. What I failed to stress in a clear fashion in
my Letter, as pointed
out by HJ, is that this symmetry restoration is equivalent to the symmetry
restoration of the Ising model due to the breakdown of the {\it mean field}
approximation. This was implicitly assumed at the end, when I used the same
notation for the critical exponent ($\beta$) controlling the divergence of the
order parameter $\Delta F_{{\rm eq}}$
with $\alpha_c$ as the magnetization $M$ with $T_c$ in the Ising model.
Indeed, in a previous publication \cite{GK}, I showed that
a simple field redefinition brings the system with a cubic term into a
Ginzburg-Landau model with a complicated magnetic field which vanishes at
$T_c$. No news
here, except for the fact that the quadratic term is written as
$
-{{\alpha^2}\over {18\lambda}}T_c^2\phi^2.
$
Thus, as $\alpha$ is decreased, so is the magnitude of the quadratic term,
leading to an eventual breakdown of the {\it mean field} approximation for
$\alpha < \alpha_c$.
As is well-known, the effect of incorporating fluctuations is precisely to
decrease the critical temperature from its mean field value \cite{AMIT}.
At one-loop, and in a 2d lattice, the term has the log form  with a hard
momentum cutoff described by HJ. There are also finite terms which
complicate the issue somewhat. (See also
Ref. \cite{AG}, where all these terms were computed.) Thus, by
incorporating fluctuations, the {\it effective} coarse-grained
potential (called continuum limit by HJ) would differ from its
mean field version,
in that the barrier would disappear at $\alpha_c$. (Note also
that the value of $\alpha_c$ will depend on the lattice spacing, since
the renormalization terms are lattice-space dependent!) This is the
conclusion of HJ, with which I of course agree. However, I believe
my interpretation
still remains valid, as it was based on the {\it mean field} free energy. This
is the quantity which is widely used
when describing cosmological phase transitions. In this context, the one-loop
corrected effective potential plays the same role as the mean field
Ginzburg-Landau
free energy, in that (in general)
it incorporates the effects from all fields coupled to
the scalar field, but {\it not} from the scalar field itself. Although I
focused on a real scalar field, the results are suggestive of the role of
fluctuations in destroying the ``strong'' character of the transition.
In fact, for a real scalar order parameter, ``weak'' first-order transition
is a misnomer.

I thank David Huse for his interest and comments.
This research is supported in part by the National Science Foundation
grants PHY-9453431 (Presidential Faculty Fellows award) and PHY-9204726,
and by the National Aeronautics and Space Administration grant NAGW-4270.

\end{document}